\documentclass[aps,prl,twocolumn,showpacs,preprintnumbers,amsmath,amssymb]{revtex4}    

\usepackage{graphicx}
\usepackage{dcolumn}
\usepackage{bm}
\usepackage{amsfonts}
\usepackage{amsmath,amssymb}

\newcommand {\beq} {\begin{equation}}
\newcommand {\eeq} {\end{equation}}
\newcommand {\beqa}{\begin{eqnarray}}
\newcommand {\eeqa}{\end{eqnarray}}

\begin{document}


\title{
From the planar limit to M-theory
}
 
\author{Tatsuo Azeyanagi$^{1}$}
\email{azey@physics.harvard.edu}
\author{Mitsutoshi Fujita$^{2}$}
\email{mf29@uw.edu}
\author{Masanori Hanada$^{3}$}
\email{hanada@post.kek.jp}


\affiliation{
${}^1$Center for the Fundamental Laws of Nature, Harvard University,
 Cambridge, Massachusetts 02138, USA\\
${}^2$ Department of Physics, University of Washington,
 Seattle, Washington 98195-1560, USA\\
${}^3$ KEK Theory Center, High Energy Accelerator Research Organization (KEK), 
Tsukuba 305-0801, Japan
		 }

\date{\today; preprint: KEK-TH-1579
}


\begin{abstract} 
The large-$N$ limit of gauge theories has been playing a crucial role in theoretical physics over the decades. 
Despite its importance, little is known outside the planar limit 
where the 't Hooft coupling $\lambda=g_{YM}^2N$ is fixed. 
In this Letter we consider more general large-$N$ limit --- $\lambda$ grows with $N$, e.g., $g_{YM}^2$ is fixed. 
Such a limit is important particularly in recent attempts to find the nonpertubative formulation of M-theory. 
Based on various supporting evidence, 
we propose this limit is essentially identical to the planar limit, 
in the sense the order of the large-$N$ limit and the strong coupling limit commute. 
For a wide class of large-$N$ gauge theories, 
these two limits are smoothly connected, and the analytic continuation from the planar limit is justified. 
As simple examples, we reproduce a few properties of the six-dimensional $\mathcal{N}=(2, 0)$ theory  on $S^1$
from the five-dimensional maximal super Yang-Mills theory, supporting the recent conjecture by Douglas and Lambert {\it et al.} 
that these two theories are identical.

\end{abstract}


\maketitle


\paragraph{Introduction.---}
The large-$N$ limit of gauge theories provides us with valuable insights into theoretical physics. 
So far, most of works, however, are restricted to the planar limit \cite{'tHooft:1973jz}, 
in which the 't Hooft coupling $\lambda=g_{YM}^2N$ is fixed. 
Of course there are fairly good reasons to consider the planar limit;   
for example, only the planar diagrams survive because 
the genus expansion in terms of the Feynman diagrams corresponds to the $1/N$ expansion, 
and hence the analysis is simplified drastically (see, e.g.,~\cite{'tHooft:1974hx}). 
It is also worthwhile to explore whether similar simplification takes place 
in the large-$N$ limit with $\lambda$ varied.

Recently there appeared an urgent motivation to study such a limit in the context of the AdS-CFT correspondence \cite{Maldacena:1997re}. 
In the most explored correspondence of the 4D ${\cal N}=4$ super Yang-Mills theory (SYM) and type IIB superstring theory on $AdS_5\times S^5$, 
the planar large-$N$ limit of SYM corresponds to the classical string limit on the string theory side. 
The conjecture had been tested thoroughly, because the gauge theory is controllable to some extent in the planar limit. 
It has been also conjectured that there is a CFT dual of M theory on $AdS_7\times S^4$ called
the 6D ${\cal N}=(2,0)$ theory, though its Lagrangian description is yet to be known.  
Recently conjecture by Douglas \cite{Douglas:2010iu} and 
Lambert {\it et al.} \cite{Lambert:2010iw} claims the equivalence of this theory on $S^1$ 
to the 5D SYM with maximal supersymmetry 
and that the large-$N$ limit with $g_{YM}^2$ fixed in the latter theory 
corresponds to the 11-dimensional supergravity on the M5-brane background. 
This conjecture is very important 
because compactification of the 6D ${\mathcal N}=(2,0)$ theory 
is a source of recent exciting developments like the Alday-Gaiotto-Tachikawa conjecture \cite{Alday:2009aq}, 
and hence much effort has been made to reproduce its properties from 5D SYM (see, e.g., \cite{Hosomichi:2012ek}). 
Direct calculations outside the planar limit, however, are difficult in general, and a new approach is needed. 

In this Letter we propose a novel approach to study the very strongly coupled large-$N$ limit, 
\begin{eqnarray}
N\to \infty \quad {\rm with}\quad \lambda=g_{YM}^2N \sim N^{a} \quad  (a>0).    \label{very_strong_limit}
\end{eqnarray}  
A key observation is that 
some results obtained in the planar limit are valid 
even in the very strongly coupled large-$N$ limit  \cite{Fujita:2012cf}.  
In this letter, we go one step further: we provide evidence that the 't Hooft limit and the very strongly coupled large-$N$ limit are identical, 
and for a wide class of theories the physics in the very strongly coupled large-$N$ limit can be understood as soon as one solves 
the planar limit (The precise statement will be explained in the following sections). 
As simple examples, we demonstrate that the properties of the 6D ${\mathcal N}=(2,0)$ theory
are calculated straightforwardly in our approach.  
\paragraph{Conjecture.---}\label{sec:proposal}
Now we argue the properties of the very strongly coupled large-$N$ limit.  
Our conjecture is: 
\begin{enumerate}
\item
First, the very strongly coupled large-$N$ limit is well defined. 

\item
The order of the large-$N$ limit and the strong coupling limit is commutative. 
In other words, the very strongly coupled large-$N$ limit can  be obtained by taking the leading part of the 't Hooft $1/N$ expansion 
and then sending the 't Hooft coupling to $\lambda \sim N^a\,\, (a>0)$.

\item
When there is no singularity and phase transition separating the planar and very strongly coupled regions,  
expectation values of observables in the very strongly coupled large-$N$ limit can be obtained 
by the analytic continuation from those in the 't Hooft limit. 

\end{enumerate}

To explain our conjecture more concretely, let us consider the free energy as an example. 
In the 't Hooft limit, it can be expanded as $F=\sum_{g=0}^\infty F_g(\lambda)/N^{2g-2}$. 
Conjecture 1 claims the existence of a certain $1/N$ expansion in the very strongly coupled region, 
which is not necessarily in powers of $1/N^2$ and is different from the genus expansion in general. 
For example, for 4D ${\cal N}=4$ $SU(N)$ SYM with the very strong coupling $\lambda\sim N^a\,\, (a>0)$ satisfying $1\ll\lambda\ll N$, the genus expansion $F=\sum_{g=0}^\infty F_g(\lambda)/N^{2g-2}$ is justified as the $g_s$ expansion in the dual string theory, though the coefficients $F_g(\lambda)$ is $N$ dependent (for the detail, see evidence 2 below). 
One can then rearrange it to a new $1/N$ expansion. Our conjecture claims the 
existence of a similar $1/N$ expansion even when the dual gravity description is not available. 
Conjecture 2 claims the higher genus terms of the 't Hooft expansion do not contribute to the leading part of this new $1/N$  expansion in the very strongly coupled region.
Conjecture 3 means that, as long as there is no singularity separating the 't Hooft large-$N$ limit and very strongly coupled large-$N$ limit, the expression in the latter is obtained by simply substituting $\lambda\sim N^a$ to the expression 
in the 't Hooft limit and picking up the leading order term in $1/N$.
In the following we provide some evidence supporting this conjecture. 
\vspace{-0.5cm}
\subparagraph*{\it Evidence 1: Exact results from the localization method.---}
\noindent For some quantities protected by supersymmetry,
we can confirm the validity of the analytic continuation from calculations purely on the gauge theory side. 
For example, in the Aharony, Bergman, Jafferis and Maldacena (ABJM) theory \cite{Aharony:2008ug}, we can use the localization technique to obtain the exact expression for 
the free energy \cite{Kapustin:2009kz}\cite{Drukker:2010nc, Fuji:2011km,Herzog:2010hf,Marino:2011eh} 
and circular Wilson loops \cite{Klemm:2012ii} for any $\lambda$ and $N$.
Especially in the 't Hooft limit the leading part of the free energy 
is given by $(\sqrt{2}\pi/3)N^2/\sqrt{\lambda}$ \cite{Drukker:2010nc},  
while the higher genus contributions are $(c_g\,\lambda^{2(g-1)}+\cdots)/N^{2(g-1)}$ \cite{Marino:2011eh}, 
where $c_g$ are order one constants for $g\ge1$. 
The explicit formulas show the same expression holds even at $a>0$ by substituting $\lambda \sim N^a$, and thus 
the very strongly coupled large-$N$ limit is smoothly connected from the planar limit 
(note that $\lambda\le N$ in the ABJM theory by definition).
A similar argument holds for the free energy and the 1/2 Bogomol'nyi-Prasad-Sommerfield (BPS) circular Wilson loops of the 4D $\mathcal{N}=4$ 
SYM \cite{Pestun:2007rz}. In this case, the planar dominance and the analytic continuation from the planar limit 
are valid at least at $\lambda\ll N^2$ \cite{Drukker:2000rr}. 

\vspace{-0.5cm}
\subparagraph*{\it Evidence 2: 4D ${\cal N}=4$ SYM and Wilson-'t Hooft loops.---}
\noindent As a simple but nontrivial example beyond the BPS sector, let us consider the 4D ${\cal N}=4$ $SU(N)$ SYM. 
The weakly-coupled type IIB superstring on AdS$_5\times$S$^5$ gives a good description 
when $4\pi g_s= g_{YM}^2\ll 1$ and $\alpha'/R_{AdS}^2\sim \lambda^{-1/2}\ll 1$, 
or equivalently $1\ll \lambda\ll N$ \cite{Maldacena:1997re}\footnote{
While the AdS-CFT correspondence is  well-established in the planar limit, 
one might worry if it is correct in the very strong coupling limit Eq.\eqref{very_strong_limit} with $1\ll \lambda\ll N$. 
However, numerical simulations \cite{Hanada:2009ne} and exact results obtained by using the localization method support 
the validity beyond the planar limit. 
}. 
On the gravity side, we have a double expansion with respect to $\alpha' = \ell_s^{\,2}$ $(\ell_s$ is the string length) and $g_s$.  
The expression on the gauge theory side is obtained by simply replacing $\alpha'$ and $g_s$ on the 
gravity side with proper combinations of $\lambda$ and $1/N^2$. 
Therefore, in the very strongly coupled large-$N$ limit with $0<a< 1$, 
only the leading term with respect to both $g_s$ and $\alpha'$ in the gravity side remains.   
Note that this argument holds for any operator in any large-$N$ gauge theory with a gravity dual. 

In order to go to even stronger coupling region, we can use 
the S duality and map $\lambda$ to $\tilde{\lambda}=16\pi^2 N^2/ \lambda$. 
We first set the 't Hooft coupling $\lambda$ to satisfy $1 \ll \lambda\ll N$ and 
apply the AdS-CFT correspondence to calculate physical quantities from the gravity side.  
We then map them to the very strongly coupled region $N\ll \tilde{\lambda}\ll N^2$ by the S duality. 
We stress that we just use the AdS-CFT correspondence at $1\ll\lambda\ll N$ in the usual sense 
and do not generalize it to $\lambda\gtrsim N$.   

As a concrete example, let us consider the Wilson and 't Hooft loops in the fundamental representation. 
At $1\ll\lambda\ll N$, through the AdS-CFT correspondence, 
it is calculated as the minimal surface area of the fundamental string (F1) ending 
on the Wilson loop at the AdS boundary \cite{Maldacena:1998im}. 
The 't Hooft loop is similarly evaluated except that F1 is replaced by the D string (D1).  
More concretely, on the gravity side, the Wilson loops  $\langle W(C)\rangle$ and 
the 't Hooft loops $\langle H(C)\rangle$ are calculated as  
$\log\, \langle W(C)\rangle = - \tau_{F1}S_{NG}(C)$ and $\log\, \langle H(C)\rangle = - \tau_{D1}S_{NG}(C)$,
where $S_{NG}(C)$ is the on-shell Nambu-Goto action on AdS$_5\times$S$^5$ whose boundary is given by the loop $C$. 
For simplicity we set the string length $\ell_s$ to be $\ell_s = \lambda^{-1/4}$ so that 
 the AdS radius becomes unity. In this notation, the AdS metric does not depend on the 't Hooft coupling and thus the on-shell $S_{NG}(C)$ does not 
neither. The tension of the F1 and D1 are given by 
$\tau_{F1} = \lambda^{1/2}/(2\pi)$ and  $\tau_{D1} = \tau_{F1}/g_s =  2N/\lambda^{1/2}$, 
respectively. 

In order to go to the very strongly coupled region we apply the S duality. 
It maps the Wilson loops to the 't Hooft loops and vice versa. 
In the S-dualized frame, the 't Hooft coupling and the string length are given by 
$\tilde{g}_s = 1/g_s$ and $\tilde{\ell}_s = g_s^{1/2} \ell_s$, and thus the 't Hooft coupling is 
$\tilde{\lambda} = 16\pi^2 N^2/\lambda$.   
The Wilson loops and 't Hooft loops after the S duality are given by 
\begin{eqnarray}
&&\log \langle \tilde{W}(C)\rangle = \log \langle H(C) \rangle = -\frac{2N}{\sqrt{\lambda}}\,S_{NG}(C) =  
 -\frac{\sqrt{\tilde{\lambda}}}{2\pi}\,S_{NG}(C)\,,  \nonumber\\
&& 
\log \langle \tilde{H}(C)\rangle =\log \langle W(C)\rangle= -\frac{\sqrt{\lambda}}{2\pi} \,S_{NG}(C) 
 = - \frac{2N}{\sqrt{\tilde{\lambda}}}\,S_{NG}(C)\, ,\nonumber
\end{eqnarray}
indicating that we can analytically continue the Wilson and 't Hooft loops 
in the 't Hooft large-$N$ limit to the very strongly coupled large-$N$ limit, even beyond $\lambda\sim N$. 
Note that this calculation is applicable to the Wilson and 't Hooft  loops of any shape, including non-BPS loops.  
A similar argument can be repeated in various theories with type IIB supergravity duals. 
For multiple Wilson loops, our argument is valid for the connected part of the correlation functions. 
For detail see Ref.\cite{2dYM}.

\subparagraph*{\it Evidence 3: Planar equivalence outside the planar limit.---}
\noindent Another evidence comes from the ABJM theory and its orientifold projection keeping 
a large number of supersymmetry (we call it the ABJ theory here) \cite{Aharony:2008gk} for which a clear understanding of the M-theory duals exists.  

The $U(2N)_{2k}\times U(2N)_{-2k}$ ABJM theory are dual to type IIA superstring on AdS$_4\times {\mathbb C}P^3$ 
at $1\ll\lambda\ll N^{4/5}$, and to M-theory on AdS$_4\times$ S$^7$/${\mathbb Z}_{2k}$ at $\lambda\gg N^{4/5}$. 
On the other hand the gravity dual of the $O(2N)_{2k}\times USp(2N)_{-k}$ ABJ theory is type IIA superstring on AdS$_4\times {\mathbb C}P^3/{\mathbb Z}_2$ 
and  M-theory on AdS$_4\times$ S$^7/{\mathbb D}_k$, which are obtained by taking the ${\mathbb Z}_2$ orientifold projections of the duals of the ABJM theory. 
On the gravity side, the ${\mathbb Z}_2$-invariant modes do not distinguish these two theories. 
Through the Gubser-Klebanov-Polyakov-Witten relation \cite{Gubser:1998bc}, this simple fact translates into a very nontrivial ``orientifold equivalence" \cite{Kachru:1998ys} on the gauge theory side: 
the ${\mathbb Z}_2$-invariant operators in the ABJM theory and the corresponding operators in the ABJ theory give the same correlation functions. 
Note that this equivalence holds even outside the 't Hooft limit, as long as there exists a classical gravity dual, as emphasized 
in Ref.\cite{Hanada:2011zx}. 

In the 't Hooft limit, the orientifold equivalence can be understood in various ways 
in terms of the field theory (see, e.g., \cite{Bershadsky:1998cb}).  
This equivalence is tightly related to the planarity of the large-$N$ limit, and as soon as the nonplanar 
corrections are taken into account the equivalence breaks down. 
Therefore the fact that this equivalence naturally extends to the very strongly coupled large-$N$ limit strongly suggests 
that the $1/N$ correction to the 't Hooft limit (nonplanar diagrams) is negligible also in the very strongly coupled large-$N$ limit \cite{Fujita:2012cf}. 

\vspace{-0.45cm}
\subparagraph*{\it Evidence 4: Examples without gravity duals and supersymmetry.---}
\noindent We can also find some examples that rely neither on the gravity dual nor on supersymmetry. 
The most familiar example is the Wilson's lattice gauge theory \cite{Wilson:1974sk} for $SU(N)$ pure Yang-Mills theory. 
At strong coupling, the Wilson loop behaves as $\langle W(C)\rangle\sim (g_{YM}^2N)^{-A}\times[1+O(1/N, 1/(g_{YM}^2N))]$, 
where $A$ is the minimum number of plaquettes needed to fill the loop $C$. This expansion is valid 
both in the planar and the very strongly coupled regions and clearly shows the commutativity of the strong coupling limit and  
the large-$N$ limit.  
Another example is the two-dimensional pure Yang-Mills theory \cite{Migdal:1975zg}, 
in which we can directly evaluate the free energy and Wilson loops 
to support our proposal  \cite{2dYM}.  

\vspace{-0.5cm}
\subparagraph*{\it What theories admit the analytic continuation?---}\label{sec:phase-transition}
\noindent When the string theory and M theory picture is clear, it is possible to see whether a 
given theory admits the analytic continuation. 
Let us consider the trivial vacuum of 3D $SU(N)$ maximal SYM 
(i.e., all scalars fluctuate around zero), which is dual to a stack of $N$ D2 branes. 
Because the IR fixed point is described by the ABJM theory, our question is 
whether the 3D maximal SYM and the ABJM theory are connected by the analytic continuation. 
The answer is {\it no}. 
In the very strong coupling limit, the M-theory circle transverse to the 
world volume of the D2 branes opens up and the D2-branes turn to the same numbers of 
M2 branes. These M2 branes, however, are not localized at a point in the M-theory circle, but rather 
they are smeared along the circle \cite{Itzhaki:1998dd}.  
At finite temperature, for example, the smeared M2 branes turn to a stack of coincident M2 branes described by the ABJM theory 
through the Gregory-Laflamme transition (through a shift in the moduli space at zero temperature) \cite{Gregory:1993vy}. 
Because of this, UV and IR are not connected smoothly. Indeed we can confirm it easily by using the gravity duals.

A similar argument shows that F1 must be smoothly connected to M2 branes. 
In the same manner  D4 branes and M5 branes should be smoothly connected if they are described by an identical theory, 
while NS5 branes and M5 branes cannot be smoothly connected. 

\paragraph{Application: M5 branes from 5D SYM.---}\label{sec:M5}
We apply our conjecture to the 5D $SU(N)$ maximal SYM \footnote{
Here we assume this theory is a well-defined quantum theory, as suggested in Ref. \cite{Douglas:2010iu}. 
} \footnote{
For the localization approach to this theory, see, e.g., Ref. \cite{Hosomichi:2012ek}. 
}.  
If we fix the energy scales (e.g., temperature and the distance) as we take the large-$N$ limit, 
at $1\ll\lambda\ll N^{2/3}$ we can use the dual black 4-brane picture to calculate various quantities 
in this theory \cite{Itzhaki:1998dd}\cite{Kanitscheider:2008kd}. 
Then, by analytic continuation, we can obtain the prediction for the very strongly coupled region, which is conjectured to be identical to 
the 6D $\mathcal{N}=(2, 0)$ theory on $S^1$ \cite{Douglas:2010iu}. In this section we set $\ell_s = 1$.

Let us start with the free energy. 
The Einstein frame metric for  the near horizon geometry of 
a stack of $N$ D4 branes is given by 
\begin{eqnarray}
&&ds^2_{D4} = 
\left(\frac{r}{R}\right)^{\frac{9}{8}} \left(-f(r)dt^2 
+dx^2\right) 
\nonumber\\
& &\quad\qquad
+\left(\frac{R}{r}\right)^{\frac{15}{8}}\left(\frac{dr^2}{f(r)}+r^{2}d\Omega_4^2\right),   
\label{metric_D4}
\end{eqnarray}
where $f(r) = 1-r_0^3/r^3$, $\lambda = g_{YM}^2 N = (2\pi)^2g_{s}N$, 
$R$ is given by $R^3= \lambda/(4\pi) = \pi g_s N$, and the D4-branes lie along $x$-directions. 
By using the standard method in the holographic renormalization \cite{Kanitscheider:2008kd}, the radial coordinate $r$ can be identified to the energy scale $E$
as $r=4R^3E^2$.
The nonextremality parameter $r_0$ is related to the Hawking temperature $T_H$ as
$T_H = (3/4\pi)(r_0^{\,1/2}/R^{\,3/2})$.  
By using the thermodynamic relation and 10d Newton constant 
$G_{10}=2^3\pi^6g_s^2 $, we obtain the free energy  
\begin{eqnarray}
F_{D4} = \frac{2^7 \times 5\times \pi^4 }{3^7}g_s N^3\, T_H^6\, V_4\,, \label{F_D4}
\end{eqnarray} 
where $V_4$ is  the volume of the D4 branes.  
According to our conjecture, 
this expression must hold even at $\lambda\gtrsim N^{2/3}$. 

We then show Eq.\eqref{F_D4} exactly matches with the free energy expected for the 6D $\mathcal{N}=(2, 0)$ theory. 
The $\mathcal{N}=(2, 0)$ theory on $S^1$ is dual to the M5 branes on $S^1$, whose metric is given by 
\begin{eqnarray}
ds^2_{M5} = \frac{r}{R}\left(-f(r)dt^2+dx^2 +dy^2\right) +
\frac{R^2dr^2}{r^2f(r)} +R^2d\Omega_4^2\,,  \label{metric_M5}
\end{eqnarray}
where 
$y\sim y+2\pi R_{11} $ corresponds to the M-theory circle with the radius $R_{11} =g_s$. 
Again, the energy scale is determined from the radial coordinate $r$ by $r=4R^3E^2$. 
Since the 11D Newton constant $G_{11}$ is given by $G_{11} = 2\pi R_{11} G_{10}$, we obtain the free energy of the M5-branes as 
\begin{eqnarray}
F_{M5}=  \frac{2^6\times 5\times \pi^3}{3^7}N^3\, T_H^6 \,V_5\,,   
\end{eqnarray} 
where $V_5$ is the volume of the M5 branes.    
We can easily see $F_{D4}$ and $F_{M5}$ match 
by using the relation $V_5 = (2\pi R_{11})V_4 = 2\pi g_s V_4 $.

One might have an impression that this agreement is trivial from the gravity point of view, because the M5-brane metric can be obtained by lifting up the D4-brane metric to 11D. It is, however, not true because the IIA and M-theory 
descriptions respectively are valid only at $1\ll\lambda\ll N^{2/3}$ and $\lambda\gg N^{2/3}$, and no gravity description is available at $\lambda\sim N^{2/3}$.
In order to relate them, one must turn to the gauge theory picture, which is well-defined even at the intermediate parameter region;   
the planar dominance outside the planar limit justifies such an argument.

We can also show the matching of the two-point correlators as follows. As a concrete example we consider the energy-momentum tensor $T_{yy}$ 
in the  $\mathcal{N}=(2, 0)$ theory and its dimensional reduction in 5D SYM at zero temperature. In 5D SYM, it couples to the dilaton, and the two-point function in the planar limit can be obtained 
by a standard gravity calculation \cite{Kanitscheider:2008kd}. According to our conjecture, the two-point function must have the same form at very strong coupling. 
The natural counterpart in the $\mathcal{N}=(2, 0)$ theory 
on $S^1$ is the smeared energy-momentum tensor $\int dy\, T_{yy}$. 
We can calculate the two-point function by using 11d supergravity on AdS$_7\times$S$^4$ \cite{Bastianelli:1999ab} 
with a compactified M circle.   
The result is (for the scheme-independent part)
\begin{eqnarray}
&&\int d^5 x\, e^{ipx}
\iint_0^{2\pi g_s}dy_1dy_2\,\langle T_{yy}(x,y_1)\,T_{yy}(0,y_{2})
\rangle \nonumber \\&& 
\quad\qquad=  \frac{g_s N^3}{2^3 \times 3 \times 5\times \pi^2}\,p^6\log p^2  , 
\end{eqnarray}  
which is in perfect agreement with the analytic continuation of the two-point function 
for a scalar operator in the 5D SYM in the planar limit (see Ref. \cite{Kanitscheider:2008kd} 
for the detailed calculation of it from the gravity side).  
We can show that similar agreements hold for the two-point function of the other components 
of the energy-momentum tensor. 
 
 We can also confirm the matching of the entanglement entropy through the holographic entanglement entropy 
 formula \cite{Ryu:2006bv}. The entanglement entropy for an arbitrary region $D (\subset \mathbb{R}^4)$ times $S^1$ 
 in the $\mathcal{N}=(2, 0)$ theory
 takes the same form as the one for $D$ in the 5D SYM: 
 \begin{eqnarray}
 S_{EE} = \frac{2^3\times g_s N^3}{3\pi  }\,s(D) \, .
 \end{eqnarray}
 Here $s(D)$ is a geometric factor common to the two theories,
 characterizing the minimal surface with the boundary $\partial D$ in their gravity duals 
and is independent of $g_s$ and $N$. 
Note that the agreement holds including the divergent term, 
because the UV cutoff corresponds to the same value of $r$ in Eqs.\eqref{metric_D4} and \eqref{metric_M5}. 
For F1 strings and M2 branes, we confirmed all of these agreements in the same manner.

\paragraph{Discussions.---}\label{sec:discussion}
The planar large-$N$ limit provides us with various techniques to understand nonperturbative aspects of 
quantum field theories. Our message in this Letter is that they can be straightforwardly extended to a far stronger coupling region for a wide class of large-$N$ gauge theories.  
This extension can have many applications, especially to string theory and M theory. 
Even when the analytic continuation does not work, it is possible that nice features of the planar limit survive. 
For example it would be fantastic if the integrability in the planar sector can be generalized to the M-theory limit. 
It is also interesting to extend 
the Eguchi-Kawai equivalence \cite{Eguchi:1982nm} 
to the very strongly coupled region. 
Our proposal would also be useful for large-$N$ quantum chromodynamics (QCD)\footnote{
In asymptotically free theories, including QCD, changing the 't Hooft coupling with $N$ just amounts to changing the energy scale with $N$. 
}; for example, the finite density region would be interesting 
to investigate, because the color superconductor does not exist in the 't Hooft large-$N$ limit. 
It would also be possible that a similar simplification takes place other field theories than gauge theories.   

It is natural to expect that nice properties of the $1/N$ expansion such as the uniform convergence, underlies 
this smooth connection between the 't Hooft large-$N$ limit and the very strongly coupled large-$N$ limit. 
It would be worthwhile to understand the mathematical properties of the large-$N$ gauge theories more deeply.  

From our observation (especially evidence 3), it is natural to expect that the planar dominance in the gauge theory side 
is one of the important ingredients for the classical description on the gravity side. 
It is interesting to understand emergent spacetime in gauge-gravity dualities from this viewpoint.

\paragraph*{Acknowledgement.---} 
The authors would like to thank G.~Ishiki, S.~Iso, D.~Jafferis, A.~Karch, H.~Kawai, R.~Loganayagam, T.~Okada, J.~Penedones, H.~Shimada, S.~Shimasaki, S.~Terashima, H.~Vairinhos and L.~Yaffe for stimulating discussions and comments. 
T.~A. and M.~F. are in part supported by JSPS Postdoctoral Fellowship for Research Abroad. 
T.~A. is also grateful to the Center for the Fundamental Laws of Nature at Harvard University for support. 
M.~H.  thanks the University of Washington and the University of Porto for warm hospitality during his stay.  


\end{document}